%% file: main.tex
\documentclass[11pt, a4paper, logo, twocolumn, copyright]{gensyn}
\usepackage{algorithmic}
\usepackage{graphicx}
\usepackage{textcomp}
\usepackage{xcolor}
\usepackage{hyperref}       
\usepackage{url}            
\usepackage{booktabs}       
\usepackage{nicefrac}       
\usepackage{microtype}      
\usepackage{multirow}
\usepackage{wrapfig}
\usepackage{subcaption}
\usepackage{amsthm}
\usepackage{makecell}
\newtheorem*{assumption*}{\assumptionnumber}
\newtheorem{assumption}{Assumption}
\usepackage{algorithm}
\usepackage{xspace}

\newcommand{\ourprotocol}{\texttt{CheckFree}\xspace}
\newcommand{\ourprotocolswap}{\texttt{CheckFree+}\xspace}
\input{packages}
\input{macros}

\title{All is Not Lost: LLM Recovery without Checkpoints}

\author[1,2]{Nikolay Blagoev}
\author[1]{O\u{g}uzhan Ersoy}
\author[2]{Lydia Yiyu Chen}

\affil[1]{Gensyn}
\affil[2]{Université de Neuchâtel}

\correspondingauthor{}

\reportnumber{}

\begin{abstract}
	Training LLMs on decentralized nodes or on-spot instances lowers the training cost and enables model democratization. The inevitable challenge here is the transient churns of nodes due to failures and the operator's scheduling policies, leading to losing parts of the model (some layers). The conventional approaches to recover from failures is to either use checkpointing, where periodically a copy of the entire model is sent to an additional storage, or redundant computation. 
These approaches yield significant communication and/or computation overhead even in non-failure cases and scale poorly in settings with large models. 

In this paper we propose \ourprotocol, an efficient recovery method where a failing stage (in a pipeline) is substituted by weighted averaging of the closest neighboring stages. 
In contrast to the state of the art, \ourprotocol requires no additional computation or storage. 
However, because of the nature of averaging neighbouring stages, it can only recover failures of intermediate stages. We further extend our method to \ourprotocolswap with out-of-order pipeline execution to tolerate crashes of the first and last stages.
Thanks to out-of-order pipelining, behaviour of the first and last stages are mimicked by their neighboring ones, which allows  \ourprotocolswap to recover them by copying the neighboring stages. To recover the (de-)embedding layers, \ourprotocolswap copies those layers in the neighboring stages, which requires relatively small storage overhead. We extensively evaluate our method on LLaMa models of model sizes from 124M to 1.5B with varying failure frequencies. In the case of low and medium failure rates (5-10\%), \ourprotocol and \ourprotocolswap outperform both checkpointing and redundant computation in terms of convergence wall-clock time, achieving up to \textbf{12\%} improvement over redundant computation.
Both of our proposals can be ran via our code available at: \url{https://github.com/gensyn-ai/CheckFree}.
\end{abstract}

\begin{document}

\maketitle
\input{introduction}

\input{related_work}
\input{method}

\input{evaluation}

\input{conclusion}

\bibliography{bibliography}
\bibliographystyle{plainnat}
\appendix
\input{appendices}
\end{document}

%% file: packages.tex
\usepackage{booktabs}
\usepackage{hyperref}
\usepackage{url}
\usepackage{alertmessage}
\usepackage{xspace}
\usepackage{graphicx}
\usepackage{multirow}
\usepackage{rotating}
\usepackage{tikz,lipsum}

%% file: macros.tex
\usepackage[colorinlistoftodos,textsize=tiny,textwidth=35pt]{todonotes}
\newcommand{\Comments}{1}
\newcommand{\mynote}[2]{\ifnum\Comments=1\textcolor{#1}{#2}\fi}
\newcommand{\mytodo}[2]{\ifnum\Comments=1\todo[linecolor=#1!80!black,backgroundcolor=#1,bordercolor=#1!80!black]{#2}\fi}

\ifnum\Comments=1
\paperwidth=\dimexpr \paperwidth + 50pt\relax
\oddsidemargin=\dimexpr\oddsidemargin + 25pt\relax
\evensidemargin=\dimexpr\evensidemargin + 25pt\relax
\marginparwidth=\dimexpr \marginparwidth + 25pt\relax
\fi

\usepackage{array}
\newcolumntype{P}{>{\centering\arraybackslash}p{2.5cm}}
\newcolumntype{M}{>{\centering\arraybackslash\footnotesize}m{.78cm}}
\newcolumntype{S}{>{\centering\arraybackslash\tiny}m{2cm}}

\newtcbtheorem[auto counter,number within=section]{obs}%
{Observation}{fonttitle=\bfseries\upshape, fontupper=\slshape,
	arc=0mm, colback=cyan!5!white,colframe=cyan!75!white}{theorem}

\newtcbtheorem[auto counter,number within=section]{ins}%
{Insight}{fonttitle=\bfseries\upshape, fontupper=\slshape,
	arc=0mm, colback=green!5!white,colframe=green!75!white}{theorem}

\usepackage{amsmath,amsfonts,bm}

\def\eqref#1{equation~\ref{#1}}

\def\1{\bm{1}}

\DeclareMathAlphabet{\mathsfit}{\encodingdefault}{\sfdefault}{m}{sl}
\SetMathAlphabet{\mathsfit}{bold}{\encodingdefault}{\sfdefault}{bx}{n}



%% file: introduction.tex
\section{Introduction}
Large Language Model (LLM) training utilizes multiple Graphics Processing Units (GPUs) to parallelize and shard the training. 
In pipeline parallelism (PP), the model is split across several GPUs/nodes where each node (usually having a single GPU) executes a {\em stage of consecutive layers} and communicates activations with the nodes of the neighbouring stages in forward and backward passes.  
Typically, PP is combined with data parallelism (DP) where several pipelines are trained in parallel on different data. In DP, several nodes concurrently perform the gradient descent of the same stage  and then synchronize their gradients at the end of an iteration with others in the same stage.

Training LLMs can take several months even on specialized high performance clusters \cite{llama3,dtfm}. 
Recent works have made use of decentralized or cloud on-spot instances (preemptible GPUs) to train models \cite{oobleck,bamboo,swarm}, as they are cheaper to rent per hour. 
However, due to the high training time, (hardware) failures during the training are inevitable~\cite{opt}.
Moreover, the number of failures is expected to be higher for training on decentralized or on-spot instances since the failures can also occur due to devices becoming unavailable~\cite{bamboo}.
\begin{figure*}[t]
    \centering
    \begin{subfigure}[t]{0.32\textwidth}
        \centering
        \includegraphics[clip, trim=0.4cm 0.5cm 0.4cm 0.4cm, width=\textwidth]{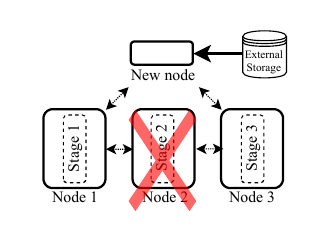}
        \caption{Checkpointing with external storage.}
    \end{subfigure}%
    ~ 
    \begin{subfigure}[t]{0.32\textwidth}
        \centering
        \includegraphics[clip, trim=0.4cm 0.5cm 0.4cm 0.4cm,width=\textwidth]{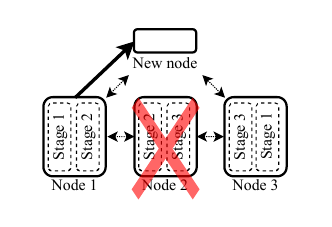}
        \caption{Redundant computation (and storage).}
    \end{subfigure}
    ~ 
    \begin{subfigure}[t]{0.32\textwidth}
        \centering
        \includegraphics[clip, trim=0.4cm 0.5cm 0.4cm 0.4cm,width=\textwidth]{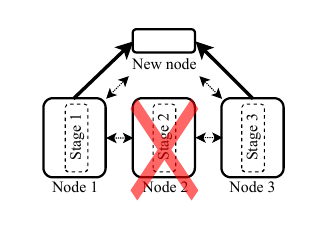}
        \caption{Our recovery method(s).}
    \end{subfigure}
    \caption{Three strategies to recover the stage failure using a new node. 
    A stage failure can happen when either each stage is only run by a single node or all the nodes of the same stage in DP fails simultaneously. 
    a) In checkpointing strategy, an external non-faulty storage is used to periodically store checkpoints. Upon a failure, a new node receives stale weights from the storage.
    b) In redundant computation (RC), each node redundantly stores and performs the computations of the subsequent stage as well~\cite{bamboo}, and in the case of failure, new node can obtain the stage from the previous one.  c) In our methods, \ourprotocol and \ourprotocolswap, we use the weights of the neighbouring stages to initialize the failed stage. \ourprotocolswap also recovers the first and last stages by copying the corresponding neighbouring stage which is trained to simulate the same behaviour. 
    }
    \label{fig:main}
\end{figure*}

{\textbf{Stage Failure}}. With data parallel training, a stage can have some or all nodes failing.
For the purposes of this work, we ignore the former scenario where at least one node remains active for a given stage, 
since failing nodes can be trivially recovered by using the alive nodes of the same stage 
as the checkpoints. For example, when new nodes join,  the most recent weights can be downloaded from a peer in the same stage, due to the data parallelism within the stage~\cite{oobleck}.
We thus focus on entire stage failures, i.e. when all nodes within a stage become unavailable.
This could happen when either (i) there is no DP and each stage is only run by a single node or (ii) all the nodes of the same stage fails simultaneously. 
The latter case is plausible for a real deployment via spot instances since such nodes of the same stage would be rented from the same geographic location (to minimize the DP communication overhead)~\cite{dtfm}, and in the case of high demand on that location, all spot instances could become unavailable simultaneously~\cite{bamboo}.

\begin{table*}[t]
    
    \begin{center}
    \caption{Comparison of failure recovery strategies regarding the additional costs required even in the non-failure cases. \(\mathcal{F}\) is the model, \(|S|\) is the parameter size per stage, \(\mathcal{E}\) is the embedding layer, and \(N\) are the number of nodes.}
    \label{table:comp}
    
  \begin{tabular}{|l|l|c|c|c|c|}
    \hline
         \multicolumn{2}{|c|}{Recovery method}  & \makecell{Checkpoint \\ \cite{gemini}} & \makecell{RC\\ \cite{bamboo}} & \ourprotocol & \ourprotocolswap \\
        \hline
        \hline
        \multirow{4}{*}{Cost} & Memory  & \(O(|\mathcal{F}|)\) & \(O(|\mathcal{F}|)\)  & 0 & \(O(|\mathcal{E}|)\) \\
        \cline{2-6}
        & Communication  & \(O(|\mathcal{F}|)\) & \(O(|\mathcal{F}|)\) & \(0\) & \(O(|\mathcal{E}|)\) \\
        \cline{2-6}
        & Computational  & 0 & Forward pass & 0 & 0 \\
        \cline{2-6}
        & Non-faulty storage & Yes & No & No & No\\
        \cline{1-6}
        \multirow{3}{*}{Recovery} & non-consecutive stages  & Yes & Yes &  Yes & Yes \\
        \cline{2-6}
         & consecutive stages  & Yes & No &  No & No \\
        \cline{2-6}
         & first and last stages & Yes & Yes &  No & Yes \\
        \cline{1-6}
  \end{tabular}
  
  \end{center}
\end{table*}

 To deal with stage failures, conventional approaches make use of {\em checkpointing} - periodically saving the model state to a non-volatile storage, which should remain reachable throughout the training~\cite{gemini}. When a failure occurs, the state of the model is reverted to the previously stored one, thus losing several GPU hours of training time. A careful tradeoff is chosen between checkpointing frequency and training throughput, as checkpointing too frequently to a remote storage can slow down the training time~\cite{gemini}. For example, LLaMa 70B with optimizer state requires roughly 700GBs when serialized and even on high-bandwidth CPU-to-GPU bus and 500Mb/s connection, it would require upwards of 20 minutes per checkpoint, thus impacting training at high checkpointing frequencies. 
Alternatively to checkpointing, Bamboo~\cite{bamboo} uses redundant computation across the pipeline to recover failures where each node stores and does redundant computation for the following stage (in addition to their assigned stage).

Both recovery methods require significant overhead either in storage, communication or computation.

In this work, we propose an efficient recovery method, \ourprotocol, that can recover even when an entire stage (consisting of multiple layers) has failed without the need for checkpointing or redundant computations. To achieve this, we exploit layer stacking \cite{stackingnesterov} and LLMs' natural resilience to layer omission \cite{layerskip}. When a stage has crashed, we reinitialize it with a weighted average of its two neighbouring stages, on a new node. While this proves a lossy recovery mechanism, we extensively demonstrate that the benefit of cheaper recovery outweighs the negatives of slower convergence. Still, this strategy fails to recover the first and last stages without suffering from significant convergence drop. 
For this case, we introduce \ourprotocolswap which utilizes the idea of out of order PP execution \cite{skippipe}, such that the second and second to last stages learn the behaviour of the first and last stage respectively. A visualization of the different approaches is presented in Fig.~\ref{fig:main}, and their comparison is given in Table~\ref{table:comp}.

\par The rest of the paper is structured as follows. We first present relevant background on fault tolerance and checkpointing. We then present our recovery method, with theoretical justification of its stability. We further extend our work with a novel use of out of order pipelining for knowledge encoding. Lastly, we extensively evaluate our solution across various models and failure ratios and demonstrate superior performance of our method, which improves training time by over \textbf{12\%} relative to redundant computation in low failure frequencies.

%% file: related_work.tex
\section{Related Work}

\paragraph{Failure Recovery} 
As the models are growing and training process becoming ever more expensive, fault-tolerant distributed LLM training has been receiving greater recognition \cite{swarm,bamboo,oobleck,faulttolerant}. SWARM~\cite{swarm} and DMoE~\cite{dmoe} studied efficient recovery in PP training in the presence of failures. Recovery of weights and optimizer states has typically been performed through checkpointing~\cite{gemini}, where periodically the weights and optimizer states are stored in some non-faulty remote storage. In case of a failure, new coming nodes download the corresponding weights from the remote storage. The training proceeds from the previous checkpoint. 

In decentralized training via spot instances, churn rates are even higher and there may not be an available non-faulty storage to save the checkpoints. 

Alternatively, Bamboo~\cite{bamboo} makes use of redundant computation, where each node computes the forward pass for itself and redundantly for the next node in the pipeline. In the case of a failure, the previous node can immediately continue the training with the redundant weights. However, such mechanism can prove too costly for large models, as each node performs redundant forward pass~\cite{oobleck}. Instead, Oobleck~\cite{oobleck} uses of the data parallel replication to recover from failures, where new nodes can get the weights from any of the nodes servicing the same weights as it. While efficient, such a recovery mechanism fails when an entire stage is gone.

\paragraph{Layer Stacking} 
Layer (or model) stacking has emerged as an efficient method for training LLMs by starting from a small model and periodically initializing new weights \cite{stackingnesterov,stacking,Midas}. Across commonly used strategies such as copying a layer to increase the depth and random initialization, it is shown that  copying the last layer provides the biggest speed up and has the best over all performance, even outperforming training the full model~\cite{stacking}. In \cite{stackingnesterov}, authors further demonstrated that such a method of initialization approximates accelerated Nesterov gradient descent. Finally, in \cite{Midas}, authors showed the resemblance of models trained with stacking to looped LLMs \cite{looped} and further found that stacking new layers in the middle greatly improves the model's reasoning for downstream tasks. 

\paragraph{Layer Omission}  LayerSkip showed that during training layers can be skipped to improve early exit inference and to speed up training \cite{layerskip}. SkipPipe further demonstrated that arbitrary layer skipping and out of order execution are a viable methods to speed up distributed training \cite{skippipe}. Both of these are reminiscent of earlier work on Stochastic Depth, which has been shown to work as gradient noise regularization \cite{residualcon}. Moreover, the results of \cite{growandprune,everyparam} showed that models can converge even with layers removed/skipped during training. This implies a certain redundancy between neighbouring layers of LLMs.

%% file: method.tex
\section{Problem Statement}

We assume a LLM model \(\mathcal{F}\), composed of an embedding layer \(\mathcal{E}\), several layers with residual connections \(f_1,\ldots,f_L\), and a deembedding layer \(\mathcal{E}^{-1}\), \(\mathcal{F} = \mathcal{E}^{-1} \circ (I + f_L) \circ \ldots \circ (I + f_2) \circ (I + f_1) \circ \mathcal{E} \), where \(\circ\) is a composing operation and \(I\) is the identity matrix. The weights of each layer \(l\in[1,L]\) are given by \(W_l\). The goal of the model (\(\mathcal{F}: \mathcal{X} \to  \mathcal{Y}\)) is to minimize some loss function (\(\mathbb{E}_{(x,y)\sim D} \mathcal{L}\)) over some dataset (\(D\) with distribution \(\mathcal{X}\times\mathcal{Y}\)). 

\paragraph{Distributed Training} The model is split over stages \(S_1,\ldots, S_s\), each holding a non overlapping and consecutive partition of the layers. The weights of a stage \(i \in [1, S]\) is given by \(W^S_{i}\), a shorthand for the combination of weights of the layers in that stage. Nodes serve a single stage and communicate with their peers to train the model. Communication occurs over the network and is associated with delays dependent on the bandwidth and latency between pairs of nodes. No central non-faulty storage exists to which model weights can be offloaded.

\paragraph{Failure pattern} Nodes can fail at arbitrary points. A failing node is disconnected from the training and its local progress is lost. While it may come back at some point in the future, its local state will be too outdated to be useful.

Upon failure of a node, we assume new one can be made available within a negligible amount of time (e.g. starting a new on-spot node to join the training).  We distinguish between failures where at least one node remains per stage or none. Dealing with the former type of failures is trivial as new nodes can download the weights of the stage from a node alive serving that stage~\cite{swarm, oobleck} due to data parallelism.
We henceforth focus only on recovering from failures where a stage is entirely disconnected. 

\section{Propose Recovery Method}

\subsection{Motivation}
Our recovery method draws inspiration from several key (known and experimented) observations regarding training of LLMs in the presence of layer omissions and stacking. 
    
\textbf{LLMs are resilient to layer omission.} In both training and inference time, LLMs are resilient to layer omission \cite{skippipe,layerskip,draftandverify}. This is partially attributed to the residual connections~\cite{residualcon} and suggests a degree of redundancy encoded in the layers (neighbouring layers can perform each other's functionality). 

\textbf{Layer stacking can improve LLM's performance.} 
Initializing new layers through a neighbouring one improves training efficiency as well as the model's reasoning~\cite{Midas,stacking,stackingnesterov}.

 \begin{figure}[t]
\centering
\includegraphics[clip, trim=0cm 0cm 0cm 2.41cm,width=0.9\linewidth]{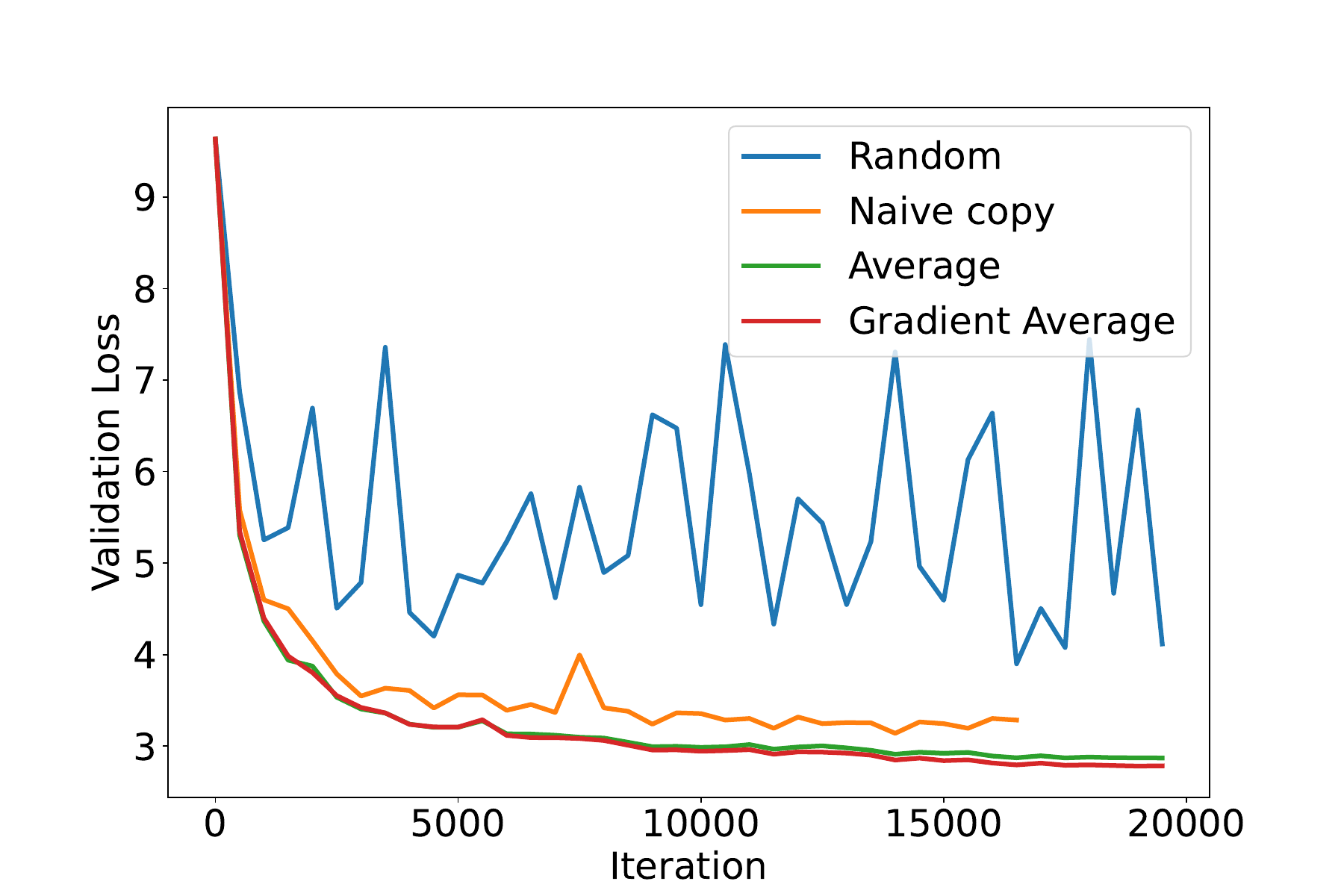}
\caption{Varying reinitialization strategies for failed stages of a LLaMa 500M model.}
\label{fig:initializat}
\end{figure}

Based on these observations, we consider recovering a failed stage with a combination of neighboring ones.
In Fig.~\ref{fig:initializat}, we present our early results on comparing four strategies: random (randomly initializing the stage), copy (copying the previous stage), and average (initializing by averaging of neighbouring stages), gradient average (initializing by weighted averaging of neighbouring stages wrt their gradients). Here, we train a LLaMa 500M model on the OpenWebText  dataset~\cite{owbtx} with a failure probability of 16\% for any stage (but those holding embedding and de-embedding).
It can be seen that though copying a neighbouring stage is significantly better than random reinitialization (of the failed stage), averaging methods significantly outperform them.
In the following sections, we explain how we choose the weights of the averaging (for \ourprotocol) as well as how we can recover the first and last stages as they have only a single neighbour (via \ourprotocolswap).

\subsection{\ourprotocol: Memoryless Recovery}

In the case of no checkpointing or redundant computation, if a stage failure happens, the weights of the failed stages are lost. Our method does not directly recover the exact weights of the failed stage, but recovers (maintains) the model's performance by replacing the failed stages with a combination of the remaining ones. 
Specifically, by using the observations given in the previous section, we replace the failed stage with the weighted average of the weights of its neighbouring stages, i.e.
\[W^S_{i} = \frac{\omega_{i-1}*W^S_{i-1} + \omega_{i+1}*W^S_{i+1}}{\omega_{i-1} + \omega_{i+1}}.\]

Now, we discuss how to select the weights \(\omega_{i-1}\) and \(\omega_{i+1}\). A known approach is to copy the previous stage when initializing new weights (\(\omega_{i+1} = 0\)). However, as seen in Fig.~\ref{fig:initializat}, this proves inferior to aggregated averaging. 
A naive way of averaging would be a uniform average of the two stages, i.e. (\(\omega_{i-1} = \omega_{i+1}\)).
Such averaging does not distinguish the importance and convergence of the stages, and thereby leads to slower convergence of the overall model (as seen in Fig.~\ref{fig:initializat}).
For that reason, we use the weights of the last gradient norm of the given stage, i.e. \(\omega_{i-1} = ||\nabla W^S_{i-1}||^2\) and \(\omega_{i+1} = ||\nabla W^S_{i+1}||^2\). Conceptually, this gives more weight to stages which have not converged as much yet, thus partially offloading their functionality to the new stage. 
To achieve weighted gradient averaging, for each stage \(S_i\), the nodes need to store the gradient norm \(||\nabla W^S_{i}||^2\) and send it to the new-coming nodes. The communication and storage overhead of this is negligible, as \(||\nabla W^S_{i}||^2\) is a single scalar, which is awfully smaller than the size of the weights of the stage. Once the weights are reinitialized, to further assist the new-formed stages in diverging from their (possibly) inferior state, we scale up the starting learning rate by a small amount (in our experiments by \(1.1\)), which can change over iterations subject to the learning rate scheduler. Our solution is presented in Algorithm~\ref{alg:csf}.

\begin{algorithm}
 \caption{Recovery algorithm for stage \(i\)}
\label{alg:csf}
\begin{algorithmic}[1]
{\small
\REQUIRE{ new node assigned to stage \(i\), \(\lambda\) learning rate}
\STATE Receive \(W^S_{i-1}\) and \(\omega_{i-1} := ||\nabla W^S_{i-1}||^2\) of stage \(i-1\) 
\STATE Receive \(W^S_{i+1}\) and \(\omega_{i+1} := ||\nabla W^S_{i+1}||^2\) of stage \(i+1\) 
\STATE Initialize the weights of the failed stage and update its learning rate \(W^S_{i} \leftarrow \frac{\omega_{i-1} W^S_{i-1} +\omega_{i-1} W^S_{i+1}}{\omega_{i-1} + \omega_{i+1} }\),  \(\lambda_i \leftarrow 1.1\lambda_i\)
\STATE Continue training from the current batch

}
\end{algorithmic}
\end{algorithm}

Such a memoryless averaging approach, however, cannot recover the first and last stages, as there is no second neighbour to average with. Additionally, as identified in several other works~\cite{Midas,vitrob}, the first and last stages perform different functionality than the other stages (the first stage especially) and are critical to the performance of the model. Therefore, a naive copy of one neighbour results in a significant drop in performance. For these reasons, \ourprotocol is suitable for the recovery of intermediate stages but not for the first or last ones. In the following section we present an improvement that can recover those stages as well.

\subsection{\ourprotocolswap: First and Last Stage Recovery}

\ourprotocolswap extends \ourprotocol by also recovering the first and last stage failures. 
Here, we take inspiration from out-of-order pipeline parallelism presented in \cite{skippipe} to mimic redundant computation \cite{bamboo} without actually duplicating computation. 
Specifically, for half the microbatches, we run the stages in standard order (\(\mathcal{E},S_1,S_2...S_{s-1},S_{s},\mathcal{E}^{-1}\)) and for the other half, we swap the order of the first two and last two stages excluding the (de)embedding layers (i.e. \(\mathcal{E},S_2,S_1...S_s,S_{s-1},\mathcal{E}^{-1}\)). In this way, the neighboring stages of the first and last stages, \(S_2\) and \(S_{s-1}\), partially learn the behaviour of \(S_1\) and \(S_s\) respectively, without any additional computation, as half the time they take their places in the pipeline. For the same reason, this would result in the two layers having similar weights (within some noise range of each other), meaning they can easily be recovered from one another. Out-of-order training, however, comes with a small degradation to convergence, as noted in~\cite{skippipe}. As such, if the failure chance is almost negligible, this solution would prove slower than no checkpointing, due to the longer time needed to converge.

Finally, we explain how \ourprotocolswap recovers the embedding and deembedding (LM head) layers. While some existing works explore sharing the weights between the two \cite{sharedweights1}, thus suggesting that in case of failure of either \(\mathcal{E}\) or \(\mathcal{E}^{-1}\) the weights can be reinitialized by copying the other one's, in this work we simply send their weights to the previous and following stages. Thus in case of a failure the weights can be recovered exactly without loss of data. While this may resemble some form of decentralized checkpointing, we note that the sizes of these layers are significantly smaller compared to the other stages, thus presenting minimal communication and storage overhead.

\subsection{Convergence Analysis of \ourprotocol}

\par We base our proof on \cite{growandprune}. The model is comprised of an embedding layer \(\mathcal{E}\), deembedding layer \(\mathcal{E}^{-1}\), and a series of residual functions \(\mathcal{F} = \mathcal{E}^{-1} \circ (I +f_L) \circ ... \circ (I +f_1) \circ \mathcal{E}\). Post failure of layer \(k\)\footnote{Equivalently can be grouping of layers from \(k\) to \(k+j\), i.e. a stage}, the model is modified as \(\mathcal{F'} =  (I +f_l) \circ ... \circ (I +f_{k+1}) \circ (I +\omega_1f_{k+1} +\omega_2f_{k-1}) \circ (I +f_{k-1}) ... \circ (I +f_1) \circ \mathcal{E}\). We let \(\mathcal{F}_t\) denote the loss function at $t^{th}$ iteration. 

\begin{assumption}
The loss function \(\mathcal{L}(\mathcal{F})\) is L-smooth and convex:
$ || \mathcal{L}(\mathcal{F}_t) - \mathcal{L}(\mathcal{F}_{t+1}) || \leq L \cdot ||\mathcal{F}_t -\mathcal{F}_{t+1} ||$, $\forall t $.
\end{assumption}

\begin{assumption}
    For some \(\delta \in [0,1)\) at any timestep \(t\) the model reduction error is bounded by
    $||\mathcal{F}_t - m \odot\mathcal{F}_t ||^2 \leq \delta^2||\mathcal{F}_t||^2 $
    where \(m\) is a 0-1 vector selecting given layers of the model.
\end{assumption}
This is a common assumption in literature~\cite{everyparam,growandprune}. Given these assumptions, the convergence of a model past a failure with our recovery method is given by:
\begin{equation}
   \small
   \sum_t\mathbb{E}||\mathcal{F}' _{t-1} - \mathcal{F}_{0}||^2 \leq \mathcal{O}(\frac{1}{t}) + \sum_t2\mathbb{E}||\omega_1f_{k+1} +\omega_2f_{k-1} - f_k||^2. 
\end{equation}

\par This implies that every failure slows down convergence with the error incurred by the initialization. The details of the proof are provided in Appendix \ref{app:proof}.

%% file: evaluation.tex
\section{Evaluation}

\begin{figure*}[tp!]
\begin{center}
    \begin{subfigure}[t]{0.65\columnwidth}
        \centering
        \includegraphics[clip, trim=0cm 0cm 0cm 2.41cm,width=\textwidth]{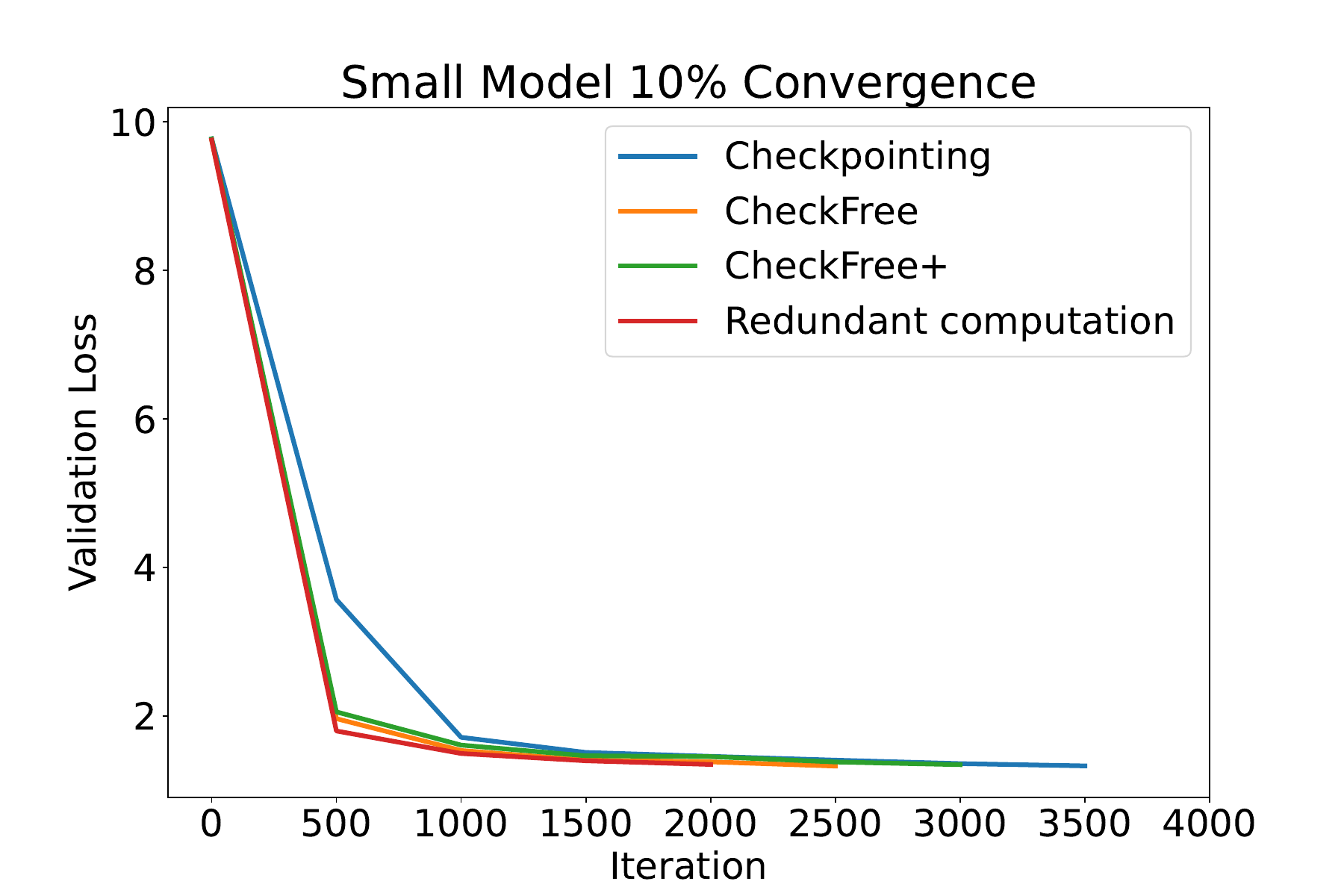}
        \caption{Small model at 10\% failure rate.}
    \end{subfigure}
    
    \begin{subfigure}[t]{0.65\columnwidth}
        \includegraphics[clip, trim=0cm 0cm 0cm 2.41cm,width=\textwidth]{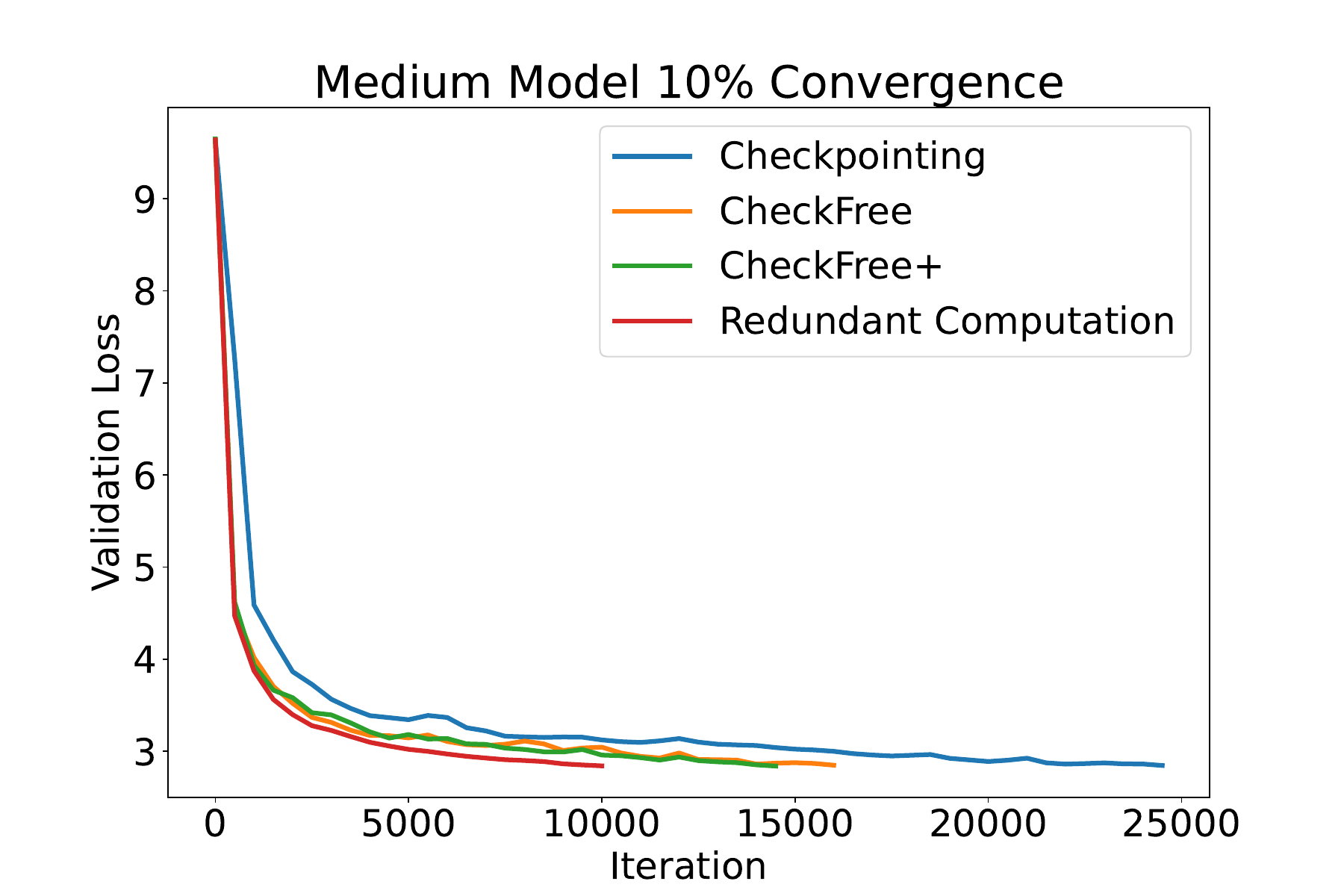}
        \caption{Medium model at 10\% failure rate.}
        \label{fig:mid-convg}
    \end{subfigure}
     \begin{subfigure}[t]{0.65\columnwidth}
        \includegraphics[clip, trim=0cm 0cm 0cm 2.41cm,width=\textwidth]{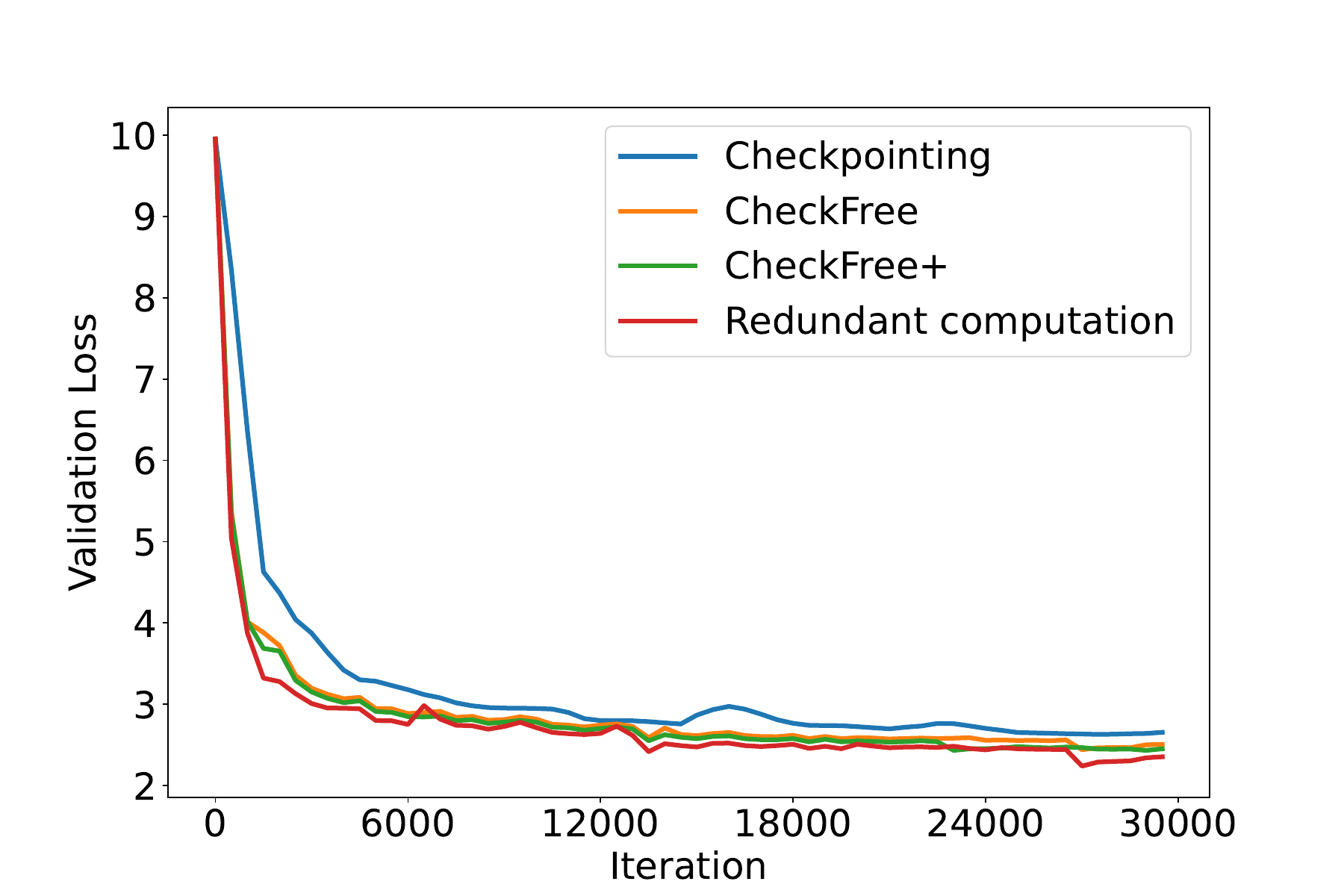}
        \caption{Large model at 16\% failure rate.}
        \label{fig:large-convg}
    \end{subfigure}
    \caption{Convergence of the models with 10\% failure rate for small and medium models \& 16\% for large ones.}
    \label{fig:convg}
\end{center}
\end{figure*}

\begin{table*}[h]
\small
\centering
    \caption{Performance of three different recovery strategies for three different failure rates until convergence for medium models (specifically, when the validation loss reaches 2.85).}
    \label{table:results}
    {\scriptsize
  \begin{tabular}{|p{2.1cm}|p{0.75cm}|p{0.75cm}|p{0.75cm}|p{0.75cm}|p{0.75cm}|p{0.75cm}|p{0.75cm}|p{0.75cm}|p{0.75cm}|p{0.75cm}|p{0.75cm}|p{0.75cm}|}
    
    \hline
         &
      \multicolumn{3}{c|}{Checkpointing} &
      \multicolumn{3}{c|}{Redundant Comp.} &
      \multicolumn{3}{c|}{\ourprotocol} & \multicolumn{3}{c|}{\ourprotocolswap} \\ \hline
      Failure rate  & 5\% & 10\% & 16\% & 5\% & 10\% & 16\% & 5\% & 10\% & 16\% & 5\% & 10\% & 16\%\\
     \hline
    \hline
    Iteration time (s) & 91.35 & 91.35 & 92.14 & 151.04 & 151.04 & 151.04 & \textbf{91.32} & \textbf{91.32} & \textbf{92.12} & \textbf{91.32} & \textbf{91.32} & \textbf{92.12} \\
    
    \hline
    Train time (h) & 558.23 & 621.67 & 634.35  & 419.56 & 419.56 & \textbf{419.56} & 367.81 & 405.87 & 562.96 & \textbf{355.13} & \textbf{367.81} & 460.6\\
    \hline  
  \end{tabular}  
  }
\end{table*}

\par In this section, we evaluate \ourprotocol and \ourprotocolswap on models ranging from 124M to 1.5B (details of each model can be found in Appendix \ref{app:models}). We demonstrate that our solutions outperform the state of the art in terms of failure recovery and does not suffer from significant decrease in convergence, despite incurring lower storage and computation cost compared to existing failure recovery strategies.

\textbf{Setup.} We perform our tests on a single H100 node with 8 GPUs. Communication delays between nodes are simulated based on realistic bandwidth and latency measurements between 5 geo-distributed locations from Google Cloud. We use the rates of 5\%, 10\%, or 16\% as the probability of a stage failure within an hour. These values are inspired from Bamboo \cite{bamboo} where they use double of these failure rates per node, while for us they represent stage failures.\footnote{These probabilities could be significantly lower. Under the assumption that there are \(k\) devices each with an independent probability of disconnecting \(p\), the chance of an entire stage failing is \(p^k\). However, in practice one would use a datacenter responsible per stage. Thus, failure of a stage is plausible if the datacenter becomes fully occupied and spot instances simultaneously become unavailable.} We expect in practice these probabilities to be significantly lower. However, this choice of higher probabilities is meant to stress test our solution in highly challenging churn rates. As we see in Fig. \ref{fig:ablation}, our solution works better at lower failure frequencies, which are more realistic.

\textbf{Baselines.}
We compare against two baselines - checkpointing and redundant computation. 
For checkpointing, we create checkpoints for the small models every 50 iterations, for the medium models every 100 iterations, and for the large models - every 40 iterations (roughly corresponding to every 3 hours, as per~\cite{bloom}). The effect of checkpointing frequency on convergence can be found in our ablation study in Appendix~\ref{app:ablation}. For redundant computation, each stage computes redundantly the next stage in the forward pass as per~\cite{bamboo}. To account for the higher memory requirement, we use half the microbatch size, but double the microbatch count, thus keeping the same batch size. Details of all experiments can be found in Appendix~\ref{app:train}.

\begin{table}[t]
\begin{center}
\scriptsize
    \caption{Perplexity of 1.5B models after 30K iterations.}
    \label{table:eval}
    {
  \begin{tabular}{|l|c|c|c|}
    \hline
     Perplexity \(\downarrow\) & No failures &  \ourprotocol & \ourprotocolswap \\
     \hline
     \hline
     OpenWebText & 16.07 & 17.59 & 17.07\\
     \hline
     Gutenberg & 13.32 & 14.58 & 13.46 \\
     \hline
     Stack Exchange & 19.85 & 21.36 & 20.86 \\
     \hline
     Arxiv & 12.25 & 13.33 & 12.46 \\
     \hline 
  \end{tabular}
  }
  \end{center}
\end{table}
\subsection{Convergence and Downstream Evaluation} 
We evaluate the models trained via our recovery method against redundant computation (which is convergence-wise equivalent to the case without failures) and checkpointing for the same iteration count. We report the validation loss over the iterations in Fig. \ref{fig:convg}. The medium and small models are trained until convergence. For the large models, due to resource constraints, we trained only for 30 thousand iterations and we chose the highest failure rate (16\%) as a means of stress testing our solution for it. Despite that, CheckFree+ converges close to the no-failure case. This can be explained by the fact that larger models typically have a higher degree of redundancy, thus allowing for such recovery mechanisms \cite{lowrank}, hinting at the potential of our solution scaling better with model sizes.

We evaluate the perplexity of the large pretrained models across four different datasets - OpenWebText \cite{owbtx}, GutenBerg books, and Stack Exchange and Arxiv as presented in~\cite{redpajama}. We present the results in Table~\ref{table:eval}. We see similar performance between a model trained with no faults and with our recovery strategies, despite the drastically different resultant weights. The models produced from our method are trained much faster wall-clock time wise, as redundant computation has a high per-iteration time.

\subsection{Throughput with Failures}
Here, we evaluate the training throughput of different recovery strategies - checkpointing, redundant computation, \ourprotocol and \ourprotocolswap. 

We perform the throughput tests for 500 iterations, simulating the failures of different stages across iterations, so that the failure patterns between tests are the same. 
In Table~\ref{table:results}, we report the iteration and train time. 
The reported train time is the wall-clock time needed to reach the "converged" iterations reported in Fig.~\ref{fig:mid-convg}. 
Specifically, for the convergence point, we use the validation loss reaching 2.85 as it saturates around that value.

We observe similar iteration time as checkpointing as the frequency of checkpointing is high enough that it does not impact iteration time. Combining the number of iterations needed to reach the convergence, checkpointing experiences a significantly higher train time due to the need to restart training.
For the same convergence point, \ourprotocolswap requires a higher number of iterations than redundant computing and a lower number than checkpointing. Due to a lower iteration time for \ourprotocolswap and \ourprotocol, the resulting training time of redundant computing is higher.
 Thus, our solution demonstrates a clear improvement in performance, being \textbf{12\%} faster at 5\% failure rate than redundant computation and significantly faster than checkpointing.  Additionally, our method has better scalability than redundant computation in low-bandwidth networks and larger models, due to the minimal overhead. Also, in case of stage failure, the recovery time of that stage is around 30 seconds. Ablation experiments can be found in Appendix~\ref{app:ablation}.

%% file: conclusion.tex
\section{Conclusion}
\par Failure recovery is critical for enabling training of LLMs on distributed and unreliable nodes. Prior works recover stage failures within a pipeline through redundant computation or checkpointing. We propose  \ourprotocol which recovers stage failures by reinitializing the failed stages with the weighted average of the neighboring stages. We further extend the protocol with \ourprotocolswap to tolerate first and last stage failures, via out of order pipelining. \ourprotocol exploits the inherent resilience of LLMs to layer omissions and efficient layer initialization techniques to recover from stage failures without requiring non-faulty storage or redundant computation. We extensively demonstrate the convergence of our method via empirical tests. We evaluate the effectiveness of \ourprotocol on LLaMa models of different sizes, showing that such lightweight recovery reduces the training time
compared to the state of the art in 5\% stage churn rate by over \textbf{12\%}. \ourprotocolswap also exhibits robust and steady convergence results for varying failure frequencies.

\textbf{Limitations and Future Work.} Our proposed solutions \ourprotocol and \ourprotocolswap cannot recover from cases of consecutive stages failing together, as there is no neighboring stages for the reinitialization strategy.
We believe that extending our methods with a lightweight checkpointing scheme will address this limitation and will explore this extension in our future work. 
Moreover, we will work on improving the convergence of our methods (especially \ourprotocolswap) in non-faulty case to reduce the number of iterations required for training.

%% file: appendices.tex
\section{Reproducibility information}
\label{app:train}
\par This section describes relevant information for reproducing our results.

\subsection{Models}
\label{app:models}
\par We train three different model sizes, all of the LLaMa family. We provide details of their hyperparameters in Table \ref{table:models}.

\begin{table}[h]
    \small
    \centering
    \caption{Model hyperparameters.}
    \label{table:models}
    {\small
    \begin{tabular}{|c|c|c|c|}
        Size & \textbf{Small} & \textbf{Medium} & \textbf{\makecell{Large \cite{layerskip}}} \\
        \hline
         Parameters & 124M & 500M & 1.5B \\
         Dim & 512 & 1024 & 2048 \\
         Heads & 8 & 16 & 16 \\
         Layers & 12 & 24  & 24 \\
         Stages & 4 &  6 & 6 \\
         Context & 512 & 1024  & 4096\\
         Learning rate & \(6\times10^{-4}\) & \(3\times10^{-4}\) & \(3\times10^{-4}\) \\
    \end{tabular}
    }
\end{table}

\subsection{Optimizers}
\par All tests were performed with the Adam optimizer with no weight decay and betas (0.9,0.999)

\subsection{Datasets}
\begin{itemize}
    \item \textbf{TinyStories} \cite{tinysstories} - Available with license CDLA-sharing-1.0. 100 texts were reserved for validation. Used for the small models.
    \item \textbf{OpenWebText} \cite{owbtx} - Available with license Creative Commons Zero. 100 texts ($\approx$1M tokens) were reserved for validation. Used for the medium models.
    \item \textbf{RedPajamas v2} \cite{redpajamav2} - Available with license Apache-2.0 license. 100 texts ($\approx$10M tokens) were reserved for validation. Used for the large models.
    
\end{itemize}
\subsection{Hardware}
\par For convergence tests we run on 2, 4, and 8 H100s for the small, medium, and large models respectively. We simulate the faults and recoveries across nodes without actual communication (other than data-parallel aggregation) to speed up iteration time. This has no difference from a test performed on geo-distributed nodes, except that runtime will be longer wall-clock wise due to the communication overhead.

\par Small models converged in roughly 8 hours, medium - in 2 days, and large models - in 2 weeks.

\par For throughput results we test on 8 H100s spawning 20 separate nodes. Communication between all 20 nodes is simulated based on realistic latency and bandwidth taken from profiling Google Cloud nodes in 5 different locations. Thus our results on throughput accurately reflect a real deployment situation on geo-distributed nodes.

\section{Ablation Studies}
\label{app:ablation}

\par Here we investigate the effects of several components individually on a model's convergence with various hyperparameters and failure rates. 

\begin{figure*}[tp!]
\centering
    \begin{subfigure}[t]{0.32\textwidth}
        \centering
        \includegraphics[clip, trim=0cm 0cm 0cm 2.41cm,width=\textwidth]{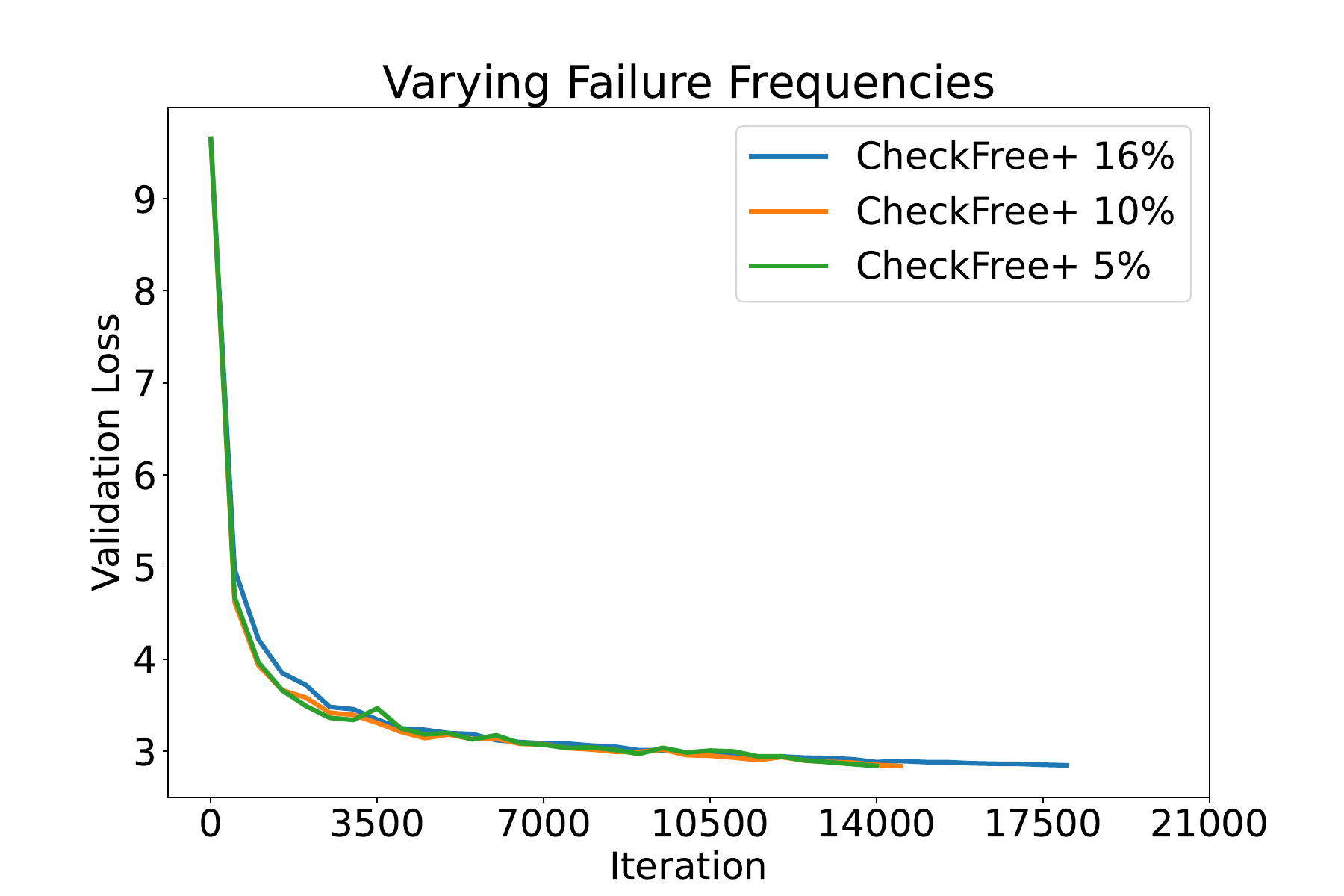}
        \caption{Varying failure frequencies}
        \label{fig:failure-freq}
    \end{subfigure}
    
    \begin{subfigure}[t]{0.32\textwidth}  
        \includegraphics[clip, trim=0cm 0cm 0cm 2.41cm,width=\textwidth]{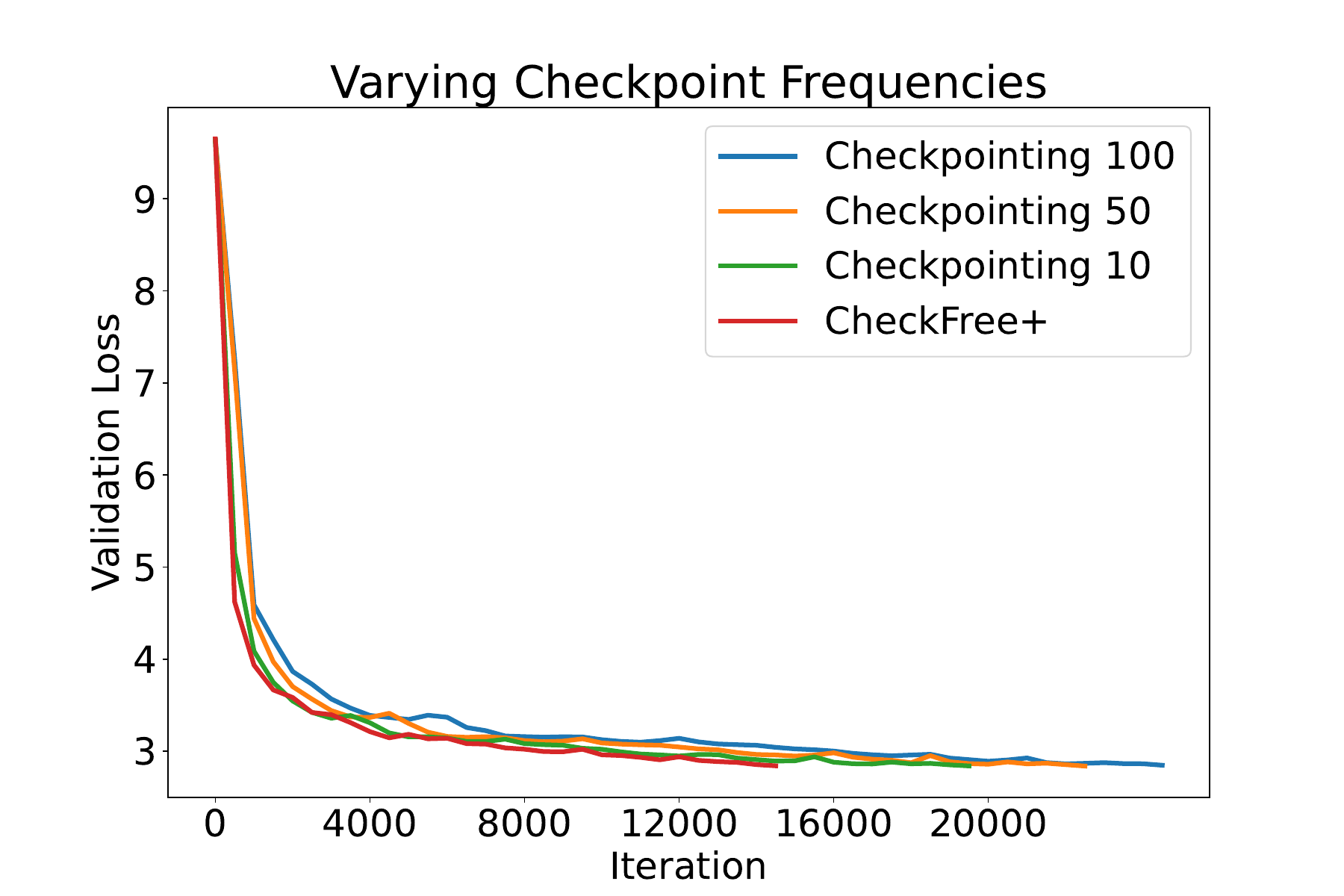}
        \caption{Varying checkpoint frequencies.}
        \label{fig:checkpoint-freq}
    \end{subfigure}
 \begin{subfigure}[t]{0.32\textwidth}
     \includegraphics[clip, trim=0cm 0cm 0cm 2.41cm,width=\columnwidth]{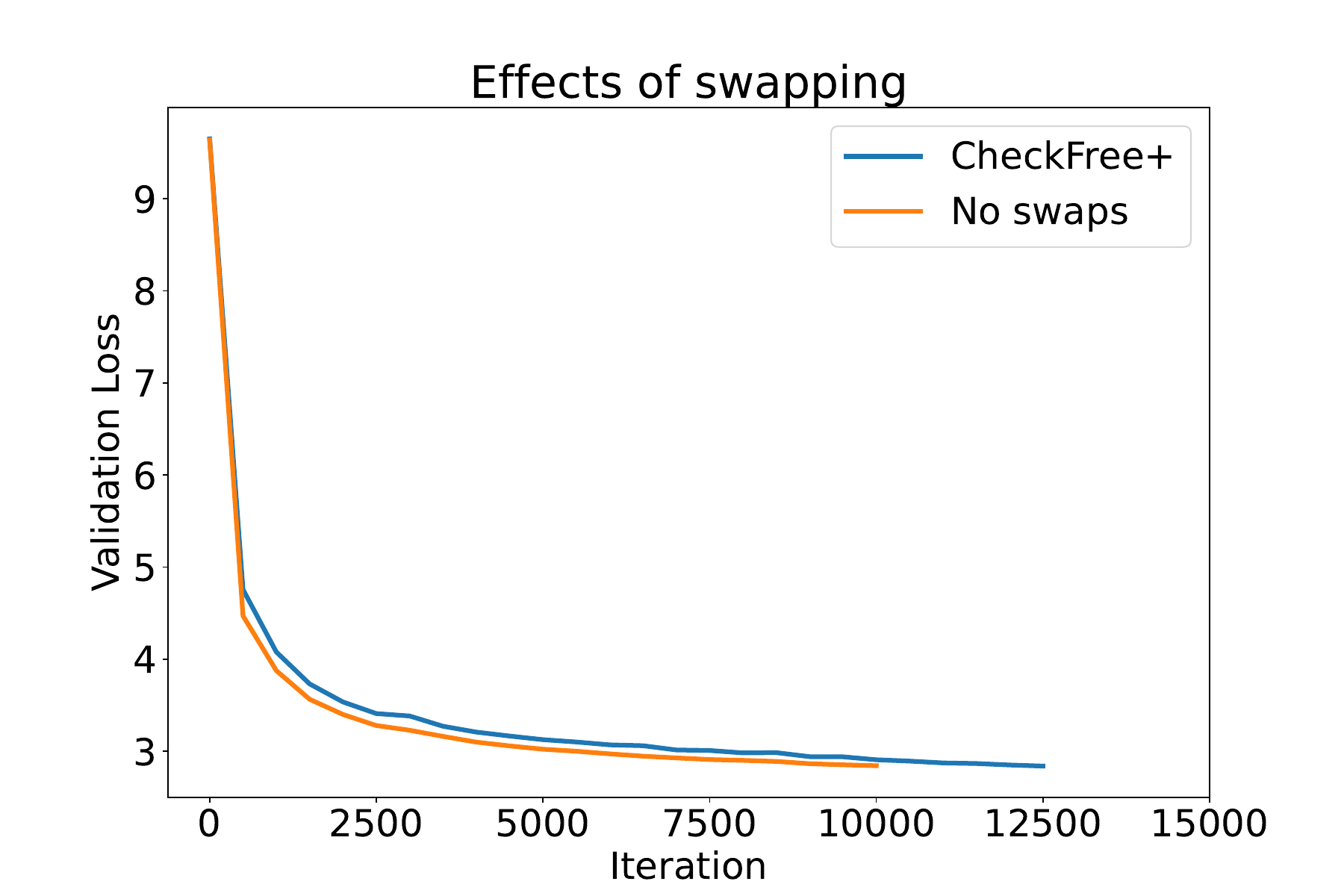}
\caption{Impact of swapping}
\label{fig:convg-swap}
    \end{subfigure}
     
    \caption{Ablation studies of  \ourprotocolswap on medium models.}
    \label{fig:ablation}
\end{figure*}

\textbf {Varying model sizes.} We empirically demonstrate the convergence of our methods on three models and dataset pairs: \textit{small} 120M LLaMa with TinyStories \cite{tinysstories}, \textit{medium} 500M LLaMa with the OpenWebText dataset \cite{owbtx}, and \textit{large} 1.5B LLaMa with the RedPyjamas dataset \cite{redpajamav2}. For the small and medium experiments we fix the crash rate at 10\%. We evaluate 4 recovery strategies: checkpointing, redundant computation, \ourprotocol and \ourprotocolswap. We present the results of the small, medium and large models in Fig.~\ref{fig:convg}. These figures demonstrate that for different model sizes, our solution is superior to checkpointing convergence-wise. Note that the figures plot over iteration count, not wall-clock time. As such, despite redundant computation converging faster in terms of iterations, they incur a high overhead, which decreases their throughputs.

While our solution converges slower iteration-wise than redundant computation, it significantly outperforms it wall-clock-wise.

\textbf{Varying failure frequency.}
\label{sec:failure-freq} One important question to consider is how our method scales with different failure frequencies. Here, we investigate this by repeating the tests on the medium (500M) model in 3 settings - 5\%, 10\%, and 16\% stage failure chance. We summarize the results in Fig.~\ref{fig:failure-freq}. 
As expected, \ourprotocolswap performs the best in low-failure settings. Yet, it can be seen that the performance (validation loss) slightly degrades even when the failure rate is tripled, which demonstrates the robustness of our recovery method.

\textbf{Checkpointing frequency.}
 Our initial comparison against checkpointing assumes a checkpoint roughly every 100 iterations (corresponding to around every 3 hours~\cite{bloom}). A higher checkpointing rate is expected to yield better convergence results (due to the significantly smaller loss of training) but incurs higher communication overhead to send the model weights to the external non-faulty storage. Here, we investigate the empirical tradeoff of checkpointing strategies, against the proposed \ourprotocolswap. We repeat the experiments of the medium model, comparing against 3 checkpointing frequencies - every 10, every 50, and every 100 iterations, at 10\% failure chance. The results are plotted in Fig.~\ref{fig:checkpoint-freq}. We observe that our solution outperforms even high frequency checkpointing (every 10 iterations) case, due to the need to rollback the model after every failure with checkpointing.

\textbf{Effects of swapping on convergence.} While \ourprotocolswap performs well in cases with failures, it incurs a non-negligible affect to its convergence in 0\% failure case, due to the swapping. Here we empirically quantify this effect on a medium model. We compare the convergence of a model trained with and without swapping. The results are plotted in Fig.~\ref{fig:convg-swap}. We observe a slowdown in convergence when swapping is used.

\section{Extended Proof}
\label{app:proof}

\par From the assumptions given in the main body, it follows:

\[||\mathcal{F}_t - \mathcal{F}'_t ||^2 = ||\mathcal{F}_t - m \odot\mathcal{F}_t + m \odot\mathcal{F}_t - \mathcal{F}'_t||^2\]

\par Where $m$ this time selects all layers that are the same (non-failed ones).
\begin{align*}
    ||\mathcal{F}_t - m \odot\mathcal{F}_t + m \odot\mathcal{F}_t - \mathcal{F}'_t||^2 \leq \delta_1^2||\mathcal{F}_t||^2 - \delta_2^2||\mathcal{F}'_t||^2 \\ \leq \delta (||\mathcal{F}_t||^2 - ||\mathcal{F}'_t||^2) \leq \delta(||\omega_1f_{k+1} +\omega_2f_{k-1}||^2 - ||f_k||^2)
\end{align*}

\par We can divide the training past a random failure of model \(\mathcal{F}_0\) into two parts - standard optimization with the modified model \(\mathcal{F'}\) and post-recovery reduction error:

\begin{align*}
\sum_t\mathbb{E}||\mathcal{F}'_{t-1} - \mathcal{F}_0||^2 = \sum_t\mathbb{E}||\mathcal{F}'_{t-1} - \mathcal{F}'_0 + \mathcal{F}'_0 - \mathcal{F}_0||^2 \\ \leq \sum_t\mathbb{E}||\mathcal{F}'_{t-1} - \mathcal{F}'_0||^2 + \sum_t\mathbb{E}||\mathcal{F}'_0 - \mathcal{F}_0||^2.
\end{align*}

\par The left hand side is standard optimization problem. This depends on the optimizer, but here we assume it converges inversely proportional to \(t\) - \(\mathcal{O}(\frac{1}{t})\).

\par The right hand is the error due to replacement of stages. What this tells us is that if the error is small, the convergence is not as affected (which intuitively makes sense). The right side can be rewritten as: \(||\omega_1f_{k+1} +\omega_2f_{k-1} - f_k||^2\), thus:

\[\sum_t\mathbb{E}||\mathcal{F}' _{t-1} - \mathcal{F}_{0}||^2 \leq \mathcal{O}(\frac{1}{t}) + \sum_t2\mathbb{E}||\omega_1f_{k+1} +\omega_2f_{k-1} - f_k||^2\]

\par QED.